\documentstyle[preprint,aps,epsf]{revtex}
\begin{document}
\newif\ifplot
\plottrue
%i%%%%\plotfalse
\newcommand{\RR}[1]{[#1]}
\newcommand{\intsum}{\sum \kern -15pt \int}
\newfont{\Yfont}{cmti10 scaled 2074}
\newcommand{\Y}{\hbox{{\Yfont y}\phantom.}}
\def\O{{\cal O}}
\newcommand{\bra}[1]{\left< #1 \right| }
\newcommand{\braa}[1]{\left. \left< #1 \right| \right| }
\def\Bra#1#2{{\mbox{\vphantom{$\left< #2 \right|$}}}_{#1}
\kern -2.5pt \left< #2 \right| }
\def\Braa#1#2{{\mbox{\vphantom{$\left< #2 \right|$}}}_{#1}
\kern -2.5pt \left. \left< #2 \right| \right| }
\newcommand{\ket}[1]{\left| #1 \right> }
\newcommand{\kett}[1]{\left| \left| #1 \right> \right.}
\newcommand{\scal}[2]{\left< #1 \left| \mbox{\vphantom{$\left< #1 #2 \right|$}}
\right. #2 \right> }
\def\Scal#1#2#3{{\mbox{\vphantom{$\left<#2#3\right|$}}}_{#1}
%\kern -2pt
{\left< #2 \left| \mbox{\vphantom{$\left<#2#3\right|$}}
\right. #3 \right> }}
\draft
\tightenlines
%\preprint{***}
\title{ Final state interaction effects in \boldmath ${\mu}$-capture induced two-body
decay of \boldmath${^3{\rm He}}$.
}
\author{R.Skibi\'nski$^*$, 
J.Golak$^*$, H.Wita\l a$^{*,\dagger}$, W.Gl\"ockle$^\dagger$. 
}
\address{$^{*}$ Institute of Physics, Jagiellonian University,
PL- 30059 Cracow, Poland}
\address{
$^\dagger$Institut f\"ur theoretische Physik II, Ruhr-Universit\"at Bochum,
D-44780 Bochum, Germany
}

\date{\today}
\maketitle
\widetext
\begin{abstract}
The $\mu$-capture process on $^3{\rm He}$ leading to a neutron, a deuteron and 
a $\mu$-ne\-u\-tri\-no in the final state is studied. Three-nucleon Faddeev wave 
functions for the initial $^3{\rm He}$ bound and the final neutron-deuteron 
scattering states are calculated using the BonnB and Paris nucleon-nucleon
potentials. The nuclear weak current operator is restricted to impulse 
approximation. Large effects on the decay rates of the final state 
interaction are found.
The comparison to recent experimental data shows that the inclusion of  
final
state interactions drastically improves the description of the data.
\end{abstract}

\pacs{ PACS numbers: }
\pagebreak
%\tableofcontents
%\pagebreak
\narrowtext

\section{Introduction.}

Recently considerable progress has been achieved in the calculations of three nucleon (3N) continuum
states. It is now possible to generate exact 3N bound
\cite{nogg1} and scattering \cite{gloc2} states for any realistic nucleon -
nucleon interaction even with inclusion of a three-nucleon force (3NF). This
opens now the possibility to study the interaction of electromagnetic and/or
weak probes with the 3N systems ($^3{\rm He}$ or $^3{\rm H}$ nuclei) without
introducing into the analysis the uncertainties due to inadequate  
approximate 3N states. Only using such exact states 
nuclear dynamics can be tested with such probes and important
information can be gained on the corresponding hadronic current operators.

Elastic and inelastic electron scattering on $^3{\rm He}$ ($^3{\rm H}$) as 
well as photodisintegration of $^3{\rm He}$ or pd capture are prominent
examples of such processes and have been studied since many years with the
hope to get insights into 3N bound state wave functions and into the hadronic
current operator. As was shown in a series of recent papers on 
electromagnetic processes \cite{ishi3}
\cite{gola4} \cite{gloc5} a very important and unavoidable ingredient for their
analysis is the exact treatment of the interaction among the
three nucleons in the continuum states. It is also important, that the      
three-nucleon wave functions should be based on realistic nuclear forces. In 
this paper we would like to study the importance of final state interactions
(FSI) between the nucleons in muon capture on $^3{\rm He}$.

There are three final channels following the capture of a negative muon on
$^3{\rm He}$:  
\begin{eqnarray}
\mu^- \;+\; ^3{\rm He}&\;\rightarrow\;&^3{\rm H} \;+\; \nu_{\mu}, \nonumber \\
\mu^- \;+\; ^3{\rm He}&\;\rightarrow\;&{\rm d} \;+\; {\rm n} \;+\; \nu_{\mu}, \\
\mu^- \;+\; ^3{\rm He}&\;\rightarrow\;&{\rm p} \;+\; {\rm n} \;+\; {\rm n} \;+\; \nu_{\mu}.\nonumber
\end{eqnarray}
The capture rate for the $^3{\rm H} \; \nu_{\mu}$ channel has been extensively
studied using the elementary particle method and the impulse approximation 
\cite{phil6} \cite{cong7}. When removing the nuclear uncertainties by using accurate
three-body bound wave functions, this reaction offers the possibility to extract 
the induced pseudoscalar coupling constant with nearly the same precision as from 
the capture process 
by a free proton \cite{cong7}.

The old calculations of the total decay rates for the dn$\nu_{\mu}$ (two-body
breakup of $^3{\rm He}$) and pnn$\nu_{\mu}$ (three-body breakup of 
$^3{\rm He}$) channels performed in \cite{phil6} using the impulse 
approximation and 3N bound and scattering states generated within the Amado  
model with separable $^1{\rm S}_0$ and $^3{\rm S}_1$ two-nucleon interactions
showed that scattering effects are large. The resulting total rates were in
agreement with some of the limited and rather inaccurate experimental results
while disagreed with others \cite{phil6}.          

Recently the first measurement of energy spectra for deuteron and proton 
leaving $^3{\rm He}$ after nuclear muon capture leading to two-body and 
three-body breakup, respectively, has been reported \cite{cumm8} \cite{kuhn9}.
The partial capture rates were compared in \cite{kuhn9} to simple plane wave
impulse approximation calculations yielding fair agreement with the measured
proton energy spectrum but underpredicting the measured rate of deuteron
production by a large factor. It is the aim of the present paper to study if
the inclusion of FSI by using an exact n-d scattering state can 
account for this 
discrepancy.
In section II we 
describe our way to fully include FSI. In section III predictions for 
decay rates obtained with two realistic NN interactions: Bonn B \cite{mach10}
and Paris \cite{laco11} are shown and compared to the experimental values. We 
conclude in section IV.

\section{Theoretical formalism.}

Fig.1. depicts the kinematics of the reaction. Our basis of the muon capture on 
$^3{\rm He}$ forms the more fundamental weak-interaction capture process on
hydrogen: $\mu^{-} \;+\; {\rm p} \;\rightarrow\; {\rm n} \;+\; \nu_{\mu}.$
The initial state 
\[ \vert \;i>\;=\;\vert \;\Psi\;s_{\mu}> \;\vert \;\Psi_{^3{\rm He}}\;m\;P>\;\]
is composed of the atomic K-shell muon wave function $\vert \;\Psi\;s_{\mu}>$
with spin projection $s_{\mu}$ and the $^3{\rm He}$ state 
$\vert \;\Psi_{^3{\rm He}}\;m\;P>$ with spin projection m and four momentum
$P.$ We choose the lab system with $\vec P \;=\; 0.$ The transition leads 
to the final state 
\[ \vert \;f> \;=\; \vert \;\nu_{\mu}\;s_{\nu} >\; \vert \;\Psi^{(-)}\; P^{'} >\;. \]
where $\nu_{\mu}$ is the four momentum of the neutrino and $s_{\nu}$ its 
spin projection. The state $\vert \;\Psi^{(-)}\; P^{'} >$ is the interacting 
neutron-deuteron state with overall four momentum $P^{'}$.

The corresponding S-matrix element ${\rm S}_{fi}$ is given in first order 
perturbation theory and assuming the Fermi form for the interaction Lagrangian
by  
\begin{eqnarray}
 {\rm S}_{fi} &=& i \int d^4x < f \;\vert \;{\cal L}(x) \;\vert \;i > \;=\\
&=& i (2\pi)^4 \; \delta(P_f \;-\; P_i)\;\frac{\rm G}{\sqrt{2}} \;*\nonumber\\
&*& <\Psi^{(-)}\;P^{'} \;\vert \; {\rm I}^{\lambda}(0) \vert \;\Psi_{^3{\rm He}}\;m\; P> \;*
< \nu\;s_{\nu} \;\vert {\rm L}_{\lambda}^{\dagger}(0)\; \vert \;\Psi\;s_{\mu}> \nonumber \\
&\equiv& i (2\pi)^4 \; \delta(P_f \;-\; P_i)\;\frac{\rm G}{\sqrt{2}} \; {\rm L}_{\lambda} {\rm I}^{\lambda}\;.\nonumber 
\end{eqnarray}
The leptonic matrix element ${\rm L}_{\lambda}$ is known and is expressed 
in terms of the corresponding Dirac spinors for the neutrino $u(\vec{\nu},s_{\nu})$
and the muon $u(\vec{\mu},s_{\mu})$ by
\begin{equation}
{\rm L}_{\lambda} \;=\; \frac{1}{(2\pi)^3} \;\bar u(\vec{\nu},s_{\nu}) \gamma_{\lambda}
(1 \;-\; \gamma_5) u(\vec{\mu},s_{\mu})\;.
\end{equation}
In the nuclear matrix element neither the current operator ${\rm I}^{\lambda}(0)
$ nor the nuclear wave functions
are equally well under control. In the following we restrict ourselves to the
impulse approximation (IA) and neglect the exchange current effects. The single
nucleon current operator is parametrized in terms of      
weak-interaction formfactors $g_{i}^{V,A}$ \cite{bail12}
\begin{eqnarray}
{\rm I}^{\lambda}(0) &=& \frac{1}{(2\pi)^3} \sum_{s,s'} \sum_{\tau,\tau'} 
\int d \vec{p} \;d \vec{p'}\; a^{\dagger}(\vec p,s,\tau) \bar u(\vec p,s) \;*\nonumber\\ 
&*& \lbrace (g_1^V \;-\; 2mg_2^V) \gamma^{\lambda}\;+\; g_2^V (p\;+\;p')^{\lambda}
\;+\; g_1^A \gamma^{\lambda} \gamma^5 +g_2^A k^{\lambda} \gamma^5 \rbrace \;*\nonumber \\
&*& \tau_{-} u(\vec{p'},s') a(\vec{p'},s',\tau')\;,
\end{eqnarray}
where $a,a^{\dagger}$ are standard nucleon creation and anihilation operators,
$\tau_{-}$ is the isospin lowering operator and $k \;\equiv\; p' \;-\; p$. For the
nucleon weak formfactors we use the standard values \cite{tata13}
\begin{eqnarray}
g_1^V (k^2) &=& \frac{1}{(1\;+\;\frac{k^2}{0.71 \; [GeV]^2})^2} \;,\\
g_1^A (k^2) &=& \frac{-1.262}{(1 \;+\; \frac{k^2}{1.19 \; [GeV]^2})^2} \;,\nonumber \\
2mg_2^V (k^2) &=& \frac{-3.7}{(1\;+\;\frac{k^2}{0.71\;[GeV]^2})^2} \;,
\nonumber\\
g_2^A (k^2) &=& \frac{-2mg_1^A(k^2)}{k^2 \;+\; m_{\pi}^2} \;, \nonumber 
\end{eqnarray}
with $m_{\pi}=138.13$ MeV.

Nuclear wave functions are generated by the
nonrelativistic
Schr\"odinger equation with realistic nuclear forces. Therefore to 
be consistent, the nuclear-current should also be chosen
nonrelativistically. After a nonrelativistic reduction of (4) and introducing
standard Jacobi momenta $\vec p, \vec q$ one 
gets for ${\rm I}^{\lambda}$ defined in 
Eq.(2)
\begin{equation} \label{piecpol}
{\rm I}^{\lambda} \;=\; < \Psi^{(-)} \;\vert \; {\rm i}^{\lambda}\;\vert
\;\Psi_{^3{\rm He}}\;m> \;, 
\end{equation}
where the momentum space matrix elements of ${\rm i}^{\lambda}$ are 
\[ < \vec p \vec q \;\vert\; {\rm i}^{\lambda} (\vec Q) \;\vert\; \vec{p'} 
\vec{q'}> \;=\; \delta(\vec{p'} \;-\; \vec p) \delta(\vec{q'} \;-\; (\vec q -
\frac{2}{3} \vec Q )) {\rm I}^{\lambda}(\vec q, \vec Q) \;,\]
\begin{eqnarray}
{\rm I}^0(\vec q, \vec Q) &=& \frac{3}{(2\pi)^3}\lbrace g_1^V\;+\; g_1^A
( \frac{\vec{\sigma}\vec{\pi}}{\rm m} \;-\; \frac{\vec{\sigma} \vec{\nu}}{2{\rm m}}) \rbrace \;, \\
\vec{{\rm I}}(\vec q, \vec Q) &=& \frac{3}{(2\pi)^3}\lbrace g_1^V (\frac{\vec \pi}{\rm m}
\;-\; \frac{\vec \nu}{2{\rm m}})\;-\; \frac{\nu}{2{\rm m}}(g_1^V \;+\;
g_2^V 2{\rm m})i(\vec{\sigma} \times \hat{\nu}) \;+ \nonumber \\
&+& g_1^A \vec{\sigma} \;-\;
g_2^A \frac{1}{2{\rm m}} m_{\mu} \hat{\nu}
(\vec{\sigma} \vec{\nu}) \rbrace \;, \nonumber
\end{eqnarray}
with $\vec{\pi} \;\equiv\; \frac{2}{3} \vec{\nu} \;+\; \vec q \;;\; \vec{\nu} 
\;=\; -\vec Q$.
The treatment of the final state follows \cite{ishi3} \cite{gola4}. For the
convenience of the reader we repeat the most important steps.  

The final 
scattering state $\vert \;\Psi^{(-)} > \;=\; \vert \Psi_{nd}^{(-)} >$ is 
decomposed into Faddeev components $\vert F_1  >$
\begin{equation} \label{siedem}
\vert \Psi_{nd}^{(-)} > \;\equiv\; \frac{1}{\sqrt{3}}(1\;+\;P)\; \vert\;
F_1> \;,
\end{equation}
where $P$ is a sum of cyclical and anticyclical permutations of three nucleons.
The Faddeev component obeys the equation 
\begin{equation} \label{faddeev}
\vert \;F_1> \;=\; \vert\; \Phi_{nd} >\;+\; G_0^{(-)}t_1^{(-)}P \; \vert \;F_1>\;,
\end{equation}
and the state $\vert\; \Phi_{nd} >$ describes the free relative motion of the 
final nucleon and the deuteron. Inserting (\ref{siedem}) and (\ref{faddeev}) into
(\ref{piecpol}) yields 
\begin{equation}
{\rm I}^{\mu} \;=\; {\rm I}_{PWIAS}^{\mu} \;+\; {\rm I}_{rescatt}^{\mu} \;,
\end{equation}
where ${\rm I}_{PWIAS}^{\mu}$ (symmetrized plane wave impulse approximation)
corresponds to the case that in Eq.(6) $< \Psi^{(-)}\;\vert $ is replaced by
$\frac{1}{\sqrt{3}} <\Phi_{nd}\;\vert\;(1\;+\;P)$ and therefore no interactions
between the outgoing nucleon and the deuteron are present. For the
${\rm I}_{rescatt}^{\mu}$ term, which contains all rescattering, $< \Psi^{(-)}\;\vert $
is replaced by $\frac{1}{\sqrt{3}} <F_1 \;\vert\;Pt_1G_0(1\;+\;P)$.
The rescattering term can be written as \cite{ishi3} \cite{gola4}
\begin{equation}
{\rm I}_{rescatt}^{\mu} \;\equiv\; \frac{1}{\sqrt{3}} <\Phi_{nd}\;\vert\;P\;
\vert U^{\mu} > \;,
\end{equation}
with
\begin{equation} \label{calkowe}
\vert \;U^{\mu}> \;=\; t_1G_0(1\;+\;P){\rm i}^{\mu}(\vec Q) \;\vert\;\Psi_{^3{\rm He}}> \;+
\; t_1G_0P \; \vert \;U^{\mu}>\;.
\end{equation}
This integral equation has the same integral kernel which one also finds in the 
3N continuum \cite{gloc2}
and the same numerical methods can be used to solve it for any NN interaction. We solve
Eq.(\ref{calkowe}) in a partial wave decomposition and in momentum space. For 
details we refer to \cite{gloc2} \cite{gloc14} \cite{gloc15} \cite{wita16}.
The decay rate follows from ${\rm S}_{fi}$ in a standard way \cite{matr17} and
for the capture process with an unpolarized initial state  
and without polarization of the outgoing particles, it is 
given by
\begin{eqnarray}
d\Gamma &=& (2\pi)^2 \frac{1}{2} \frac{1}{2} \frac{(2m'\alpha)^3}{\pi}
\int d \vec{\nu} d\vec{p_d} d\vec{p_n}
\delta (P \;-\; \nu \;-\; p_d \;-\; p_n)
(2\pi)^8 \frac{{\rm G}^2}{2} \;
\sum_{m_{^3{\rm He}} \atop m_{s_{\mu}}} \sum_{m_{s_{\nu}} \atop m_n,m_d}
{\vert\; {\rm L}_{\lambda} {\rm I}^{\lambda}
\;\vert}^2 \\
&=& 8 {\pi}^2 {(2 \pi)}^2 \frac{1}{2} \frac{1}{2}\frac{(2m'\alpha)^3}{\pi}
\int E_{\nu} m_n m_d dE_{\nu} dE_d
(2\pi)^8 \frac{{\rm G}^2}{2}
\sum_{m_{^3{\rm He}} \atop m_{s_{\mu}}} \sum_{m_{s_{\nu}} \atop m_d}
\;{\vert\; {\rm L}_{\lambda} {\rm I}^{\lambda}\;\vert}^2 \;, \nonumber
\end{eqnarray}
with $P\;=\;(m_{\mu}\;+\;m_{^3{\rm He}}, \vec 0)$ in the lab. system and
$m' \;=\; \frac{{\rm m}_{\mu}\; {\rm m}_{3^{\rm He}}}{{\rm m}_{\mu}\;+\;
{\rm m}_{^3{\rm He}}} $. The factors in (13) come from averaging 
over the spin of
the initial 
particles, $(\frac{1}{2} \frac{1}{2}),$ from the states normalizations, 
$( (2\pi)^2 (2\pi)^8),$ and from 
the muon wave function (which is taken as an  
atomic K-shell wave function), $(\frac{(2m'\alpha)^3}{\pi})$,  
with
$ \alpha = \frac{1}{137.036}.$ 

\section{Results and comparison with experimental data.}

To calculate the decay rates one has to evaluate the nuclear matrix elements  
${\rm I}^{\lambda}$. The initial 3N bound state of $^3{\rm He}$ is 
always based on a 34 channel Faddeev calculation. In treating the 3N continuum we %
included all 3N partial waves with total NN angular momenta up to $j_{max}
\;=\; 2$. In order to check the convergence of the results also calculations 
restricted to 3N states with $^1S_0 \;+\; ^3S_1-^3D_1$ NN force components only and with
$j_{max}\;=\;1$ have been performed. As NN force we use the Bonn B \cite{mach10}
and Paris \cite{laco11} potentials. Let us first consider the case of a 
kinematically complete situation in the outgoing n-d channel. Since the outgoing channel
contains 3 particles and the decay originates from an initial state of zero
momentum one needs only 2 kinematical parameters to completely define the
momenta of the outgoing particles. For these 2 parameters the neutrino 
and deuteron energies ${\rm E}_{\nu}$ and ${\rm E}_d$ can be taken. The kinematically allowed
range of neutrino energies is ${\rm E}_{\nu} \in [ 0,{\rm E}_{\nu}^{max}]$
with
\begin{equation}
{\rm E}_{\nu}^{max} \;=\; \frac{(m_{\mu} \;+\; m_{^3{\rm He}})^2 \;-\; (3m_n
\;-\; \vert E_D \vert)^2}{2(m_{\mu}\;+\;m_{^3{\rm He}})} \;.  
\end{equation}
For each neutrino energy there is a range of allowed deuteron energies as  
shown in Fig.2. Each deuteron energy corresponds to a
unique angle between neutrino and deuteron momenta. In Fig.3 we show decay
rates $\frac{d^2\Gamma}{dE_d dE_{\nu}}$ for two values of the deuteron energy.
There is a drastic increase in decay rates by a factor of $\approx 200$, 
when the interactions between the neutron and the deuteron in the final state are 
included. The increase is practically independent from the deuteron and neutrino
energies. Fig.4 shows that the 
 results obtained restricting 
 to $j_{max}
\;=\;2$ NN force components might be not yet fully converged. 
Nevertheless as is clear from Fig.5 
they are  sufficient to analyse the existing data. 
The drastic effects of FSI show that any analysis
of the experimental data should be based on  realistic continuum wave functions.

Up to now only one data set for the $\mu-$capture on $^3{\rm He}$
leading to the ${\rm n} \;+\; {\rm d} \;+\; {\nu}_{\mu}$ outgoing channel exists
\cite{cumm8} \cite{kuhn9}. In order to compare our results to those data one
has to numerically integrate the decay rate at every deuteron energy over the 
unobserved neutrino energy. This leads to the results presented in Fig.5. It is
cleary seen that while the PWIAS predictions drastically underestimate the data,
the inclusion of FSI leads to a rather good agreement, which is 
fairly independent
from the NN interaction used. Also the covergence in $j_{max}$ is similar
to the kinematically complete cases as can be seen in Fig.6.

\section{Summary and conclusions.}
We calculated the decay rates for muon capture on $^3{\rm He}$ leading to the n+d+$\nu
_{\mu}$ channel. Realistic NN forces have been used (Bonn B and Paris
potentials) and both the 3N bound state and the 3N continuum scattering wave 
functions were evaluated consistently
solving the corresponding Faddeev equations. We used the most simple 
nonrelativistic single nucleon weak-current operator parametrized by the 
standard nucleon weak formfactors. Both in the kinematically complete und 
the uncomplete situations the predictions of the plane-wave impulse approximation
for the final neutron-deuteron state drastically differ by a factor of about
$\sim$ 200 from the full result including final state interaction. The 
comparison to the only existing uncomplete
data for this process reveals, that a good description of the data only results when the 
FSI is taken into account. The total failure of PWIAS in describing those 
data shows that any future analysis of this process should be performed with
bound and scattering states generated consistently from a realistic 3N 
Hamiltonian. From the theoretical point of view mesonic exchange currents 
should also be added in future analysis.

\subsection{Acknowledgements.}
This work was supported by the Polish Committee for Scientific Research under
Grant No. 2P03B03914.
The numerical
calculations were performed on the Cray T90 and T3E of the 
H\"ochstleistungsrechenzentrum in J\"ulich, Germany and on the Convex 3840 of 
the ACK-Cyfronet, Cracow, Poland under Grant No. KBN/C3840/UJ/020/1994.

\newpage
\subsection*{Figure Captions}
\begin{description}

\item{Fig.1}
{Kinematics for the ${\mu}^- \;+\; ^3{\rm He} \;\rightarrow\;
{\rm n} \;+\; {\rm d} \;+\; {\nu}_{\mu}$ reaction.}

\item{Fig.2}
{Kinematically allowed neutrino and deuteron energies.}

\item{Fig.3}
{The decay rate $\frac{d^2 \Gamma}{dE_d dE_{\nu}}$ at two values of the
deuteron energy  $E_d = 18$ MeV  and  $E_d =29$ MeV. The dashed
and dotted lines are the PWIAS predictions for the Bonn B and Paris potentials,
respectively. The solid and dashed-dotted lines are the Bonn B and Paris potential
predictions when in the final state all NN force components up to $j_{\rm max}=1$
are included.}

\item{Fig.4}
{The convergence in $j_{\rm max}$ of the decay rate $\frac{d^2 \Gamma}{
dE_d dE_{\nu}}$
at two deuteron energies as in Fig.3. The dotted, dashed and solid lines
correspond to $^1S_0 + ^3S_1-^3D_1, j_{\rm max}=1$ and $j_{\max}=2$ Bonn B
calculations, respectively.}

\item{Fig.5}
{The decay rate $\frac{d \Gamma}{dE_d}$ for the case when neutrino is
not observed. For the description of lines see Fig.3. The squares are
experimental points from \cite{cumm8} \cite{kuhn9}.}

\item{Fig.6}
{The convergence in $j_{\rm max}$ of the decay rate $\frac{d \Gamma}{
dE_d}$.
For the decription of lines see Fig.4. The squares are experimental points
as in Fig.5.}

\end{description}
\newpage
\begin{figure}
\leavevmode
\epsffile{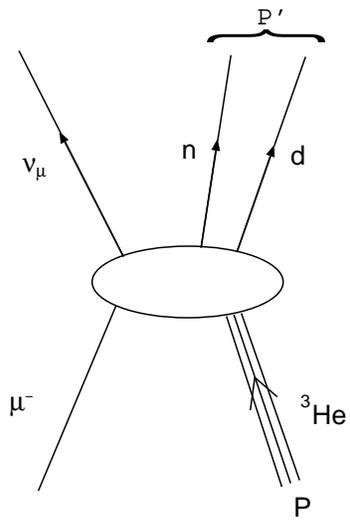}
\caption{ Kinematics for the ${\mu}^- \;+\; ^3{\rm He} \;\rightarrow\;
{\rm n} \;+\; {\rm d} \;+\; {\nu}_{\mu}$ reaction.}
\end{figure}
\begin{figure}
\leavevmode
\epsffile{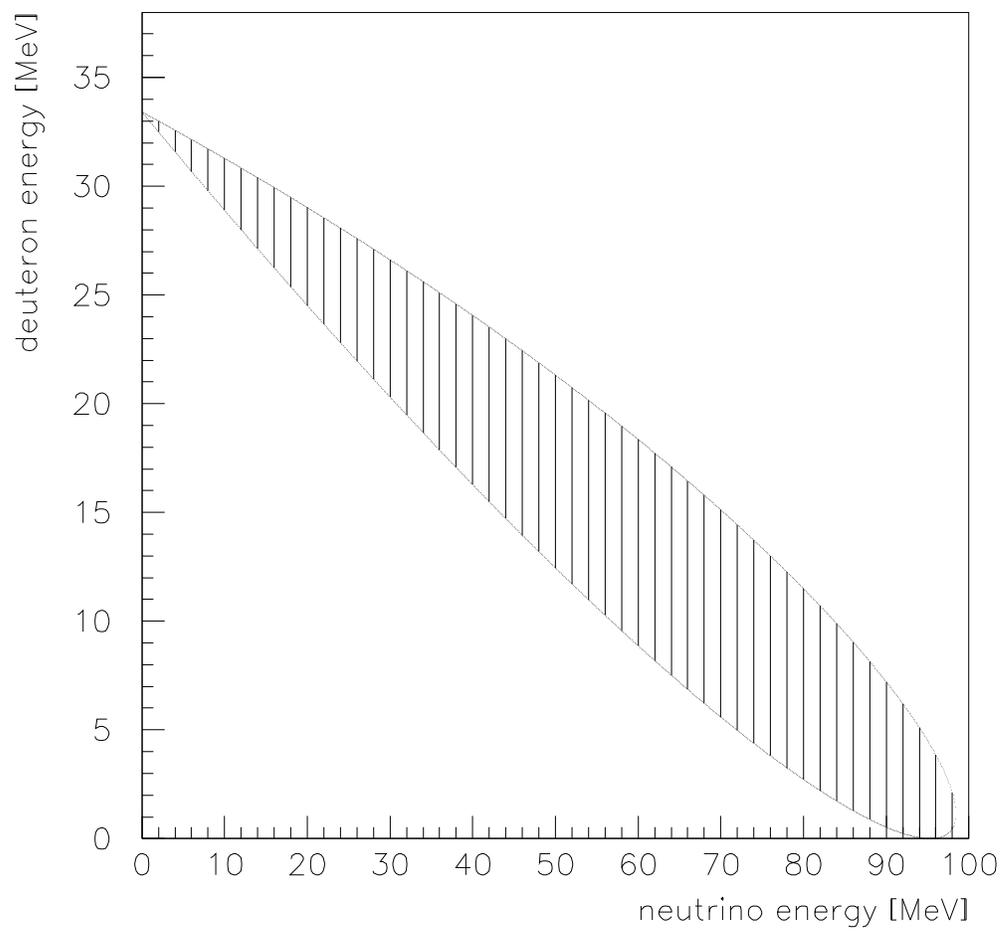}
\caption{ Kinematically allowed neutrino and deuteron energies.}
\end{figure}
\begin{figure}
\leavevmode
\epsffile{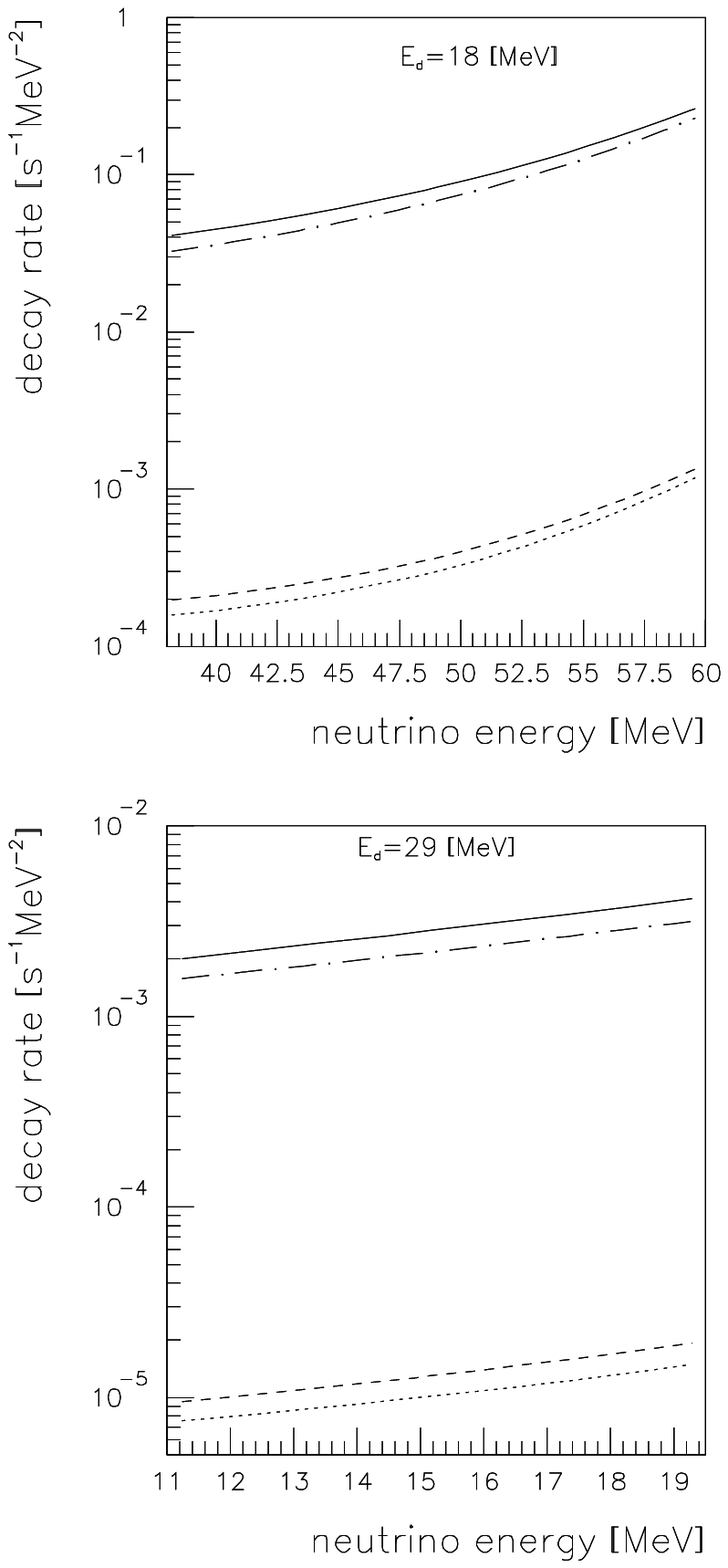}
\caption{ The decay rate $\frac{d^2 \Gamma}{dE_d dE_{\nu}}$ at two values of
the deuteron energy $E_d = 18$ MeV and  $E_d =29$ MeV. 
The dashed
and dotted lines are the PWIAS predictions for the Bonn B and Paris potentials,
respectively. The solid and dashed-dotted lines 
are the Bonn B and Paris potential
predictions when in the 
final state all NN force components up to $j_{\rm max}=1$
are included.}
\end{figure}
\begin{figure}
\leavevmode
\epsffile{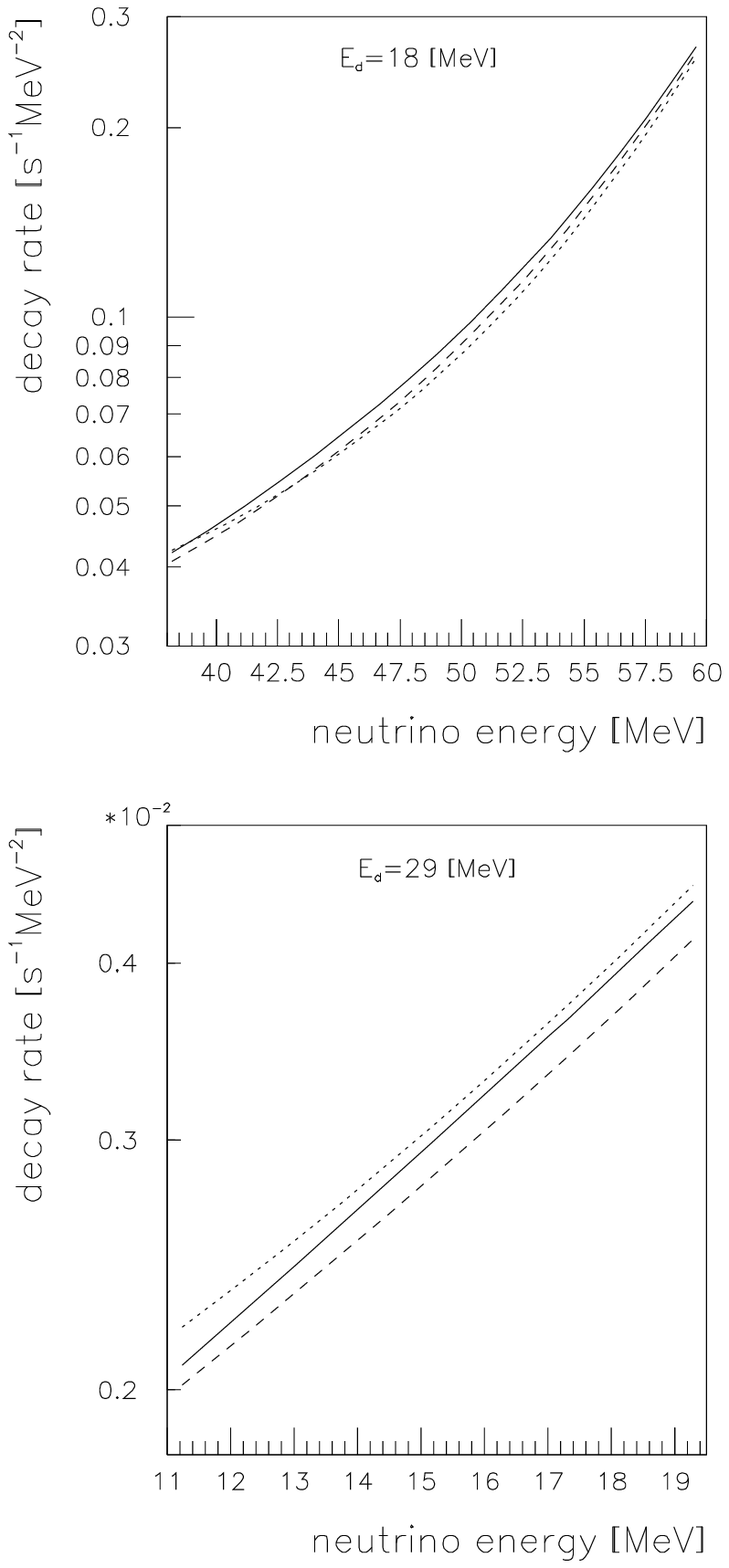}
\caption{ The convergence in $j_{\rm max}$ of the decay rate $\frac{d^2 \Gamma}{dE_d dE_{\nu}}$
at two deuteron energies as in Fig.3. The dotted, dashed and solid lines
correspond to $^1S_0 + ^3S_1-^3D_1, j_{\rm max}=1$ and $j_{\max}=2$ Bonn B 
calculations, respectively.}
\end{figure}
\begin{figure}
\leavevmode
\epsffile{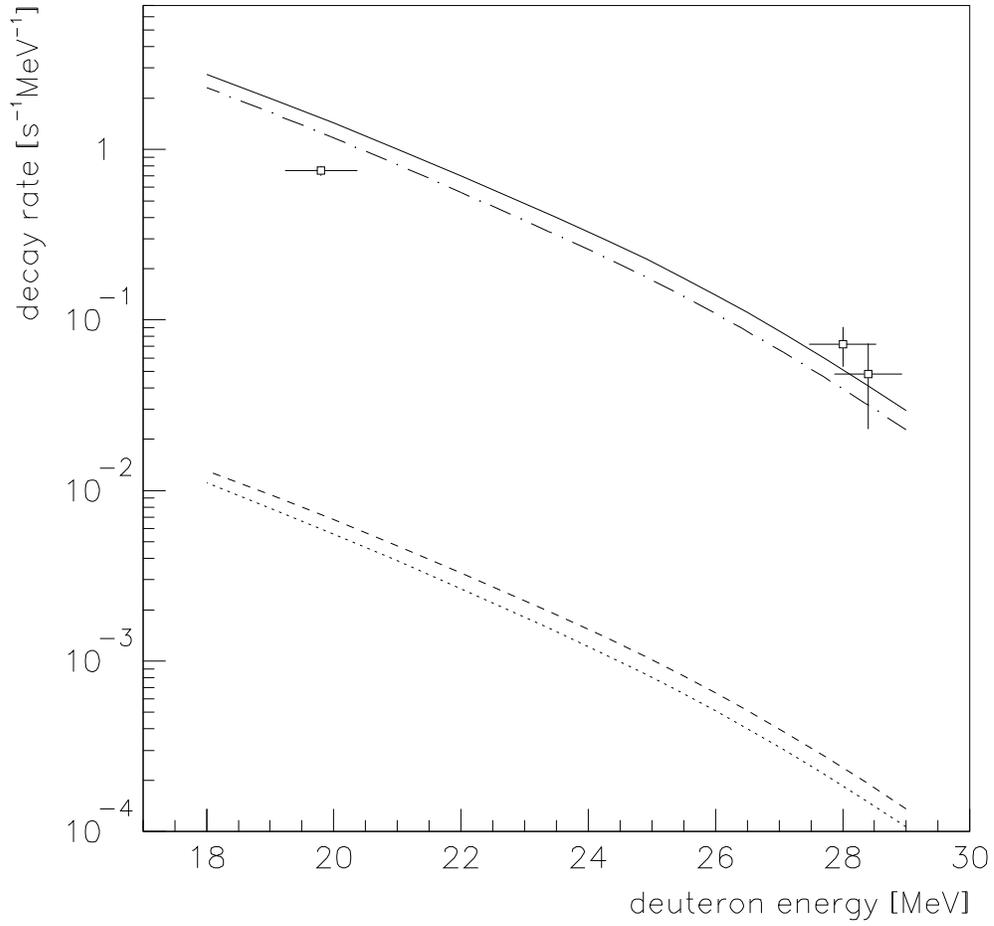}
\caption{The decay rate $\frac{d \Gamma}{dE_d}$ for the case when neutrino is 
not observed. For the description of lines see Fig.3. The squares are 
experimental points from (8) (9).}
\end{figure}
\begin{figure}
\leavevmode
\epsffile{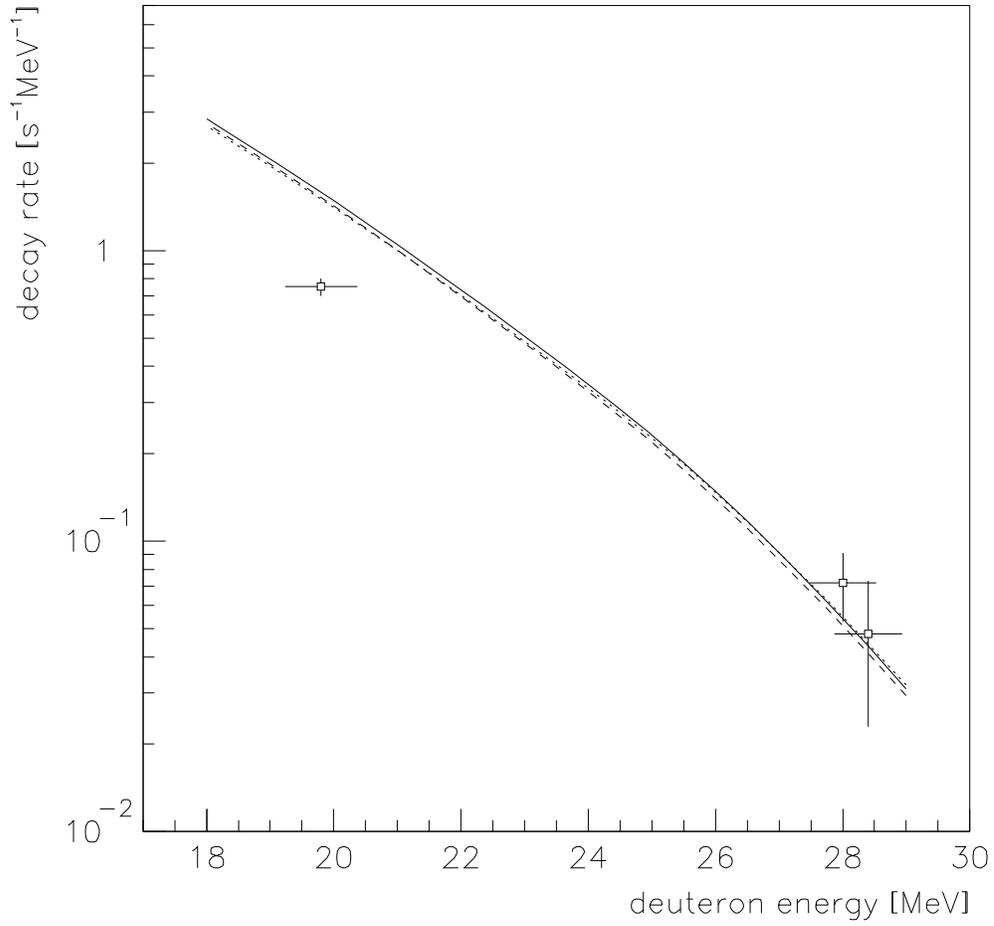}
\caption{ The convergence in $j_{\rm max}$ of the decay rate $\frac{d \Gamma}{
dE_d}$.
For the decription of lines see Fig.4. The squares are experimental points
as in Fig.5.}
\end{figure}
\end{document}